\documentclass[twocolumn, a4paper, 10pt, showpacs, prx, reprint, aps, superscriptaddress, amsmath, amssymb, amsfonts, floatfix, showkeys]{revtex4-1}

\usepackage{amsmath}
\usepackage{graphicx}
\usepackage{bm}
\usepackage[utf8]{inputenc}
\usepackage{hyperref}
\usepackage{xcolor}
\graphicspath{{figs/}{figsgaoerb/}}

\begin{document}


\title{Lattice relaxation, mirror symmetry and magnetic field effects on ultraflat bands in twisted trilayer graphene}

\author{Zewen Wu}
\affiliation{Key Laboratory of Artificial Micro- and Nano-structures of \textcolor{black}{the} Ministry of Education and School of Physics and Technology, Wuhan University, Wuhan 430072, China}

\author{Zhen Zhan}
\email{zhen.zhan@whu.edu.cn}
\affiliation{Key Laboratory of Artificial Micro- and Nano-structures of \textcolor{black}{the} Ministry of Education and School of Physics and Technology, Wuhan University, Wuhan 430072, China}

\author{Shengjun Yuan}
\email{s.yuan@whu.edu.cn}
\affiliation{Key Laboratory of Artificial Micro- and Nano-structures of \textcolor{black}{the} Ministry of Education and School of Physics and Technology, Wuhan University, Wuhan 430072, China}

\date{\today}

\begin{abstract}
Twisted graphene multilayers \textcolor{black}{exhibit} strongly correlated insulating states and superconductivity due to the presence of ultraflat bands near the charge neutral point. In this paper, the response of ultraflat bands to lattice relaxation and \textcolor{black}{a} magnetic field in twisted trilayer graphene (tTLG) with different stacking arrangements is investigated by using a full tight-binding model. We show that lattice relaxations are indispensable for understanding the electronic properties of tTLG, in particular, of tTLG in the presence of mirror symmetry. \textcolor{black}{Lattice relaxations} renormalize the quasiparticle spectrum near the Fermi energy and change the localization of higher energy flat bands. Furthermore, different from the twisted bilayer graphene, the Hofstadter butterfly spectrum can be realized at laboratory accessible strengths of magnetic field. Our work verifies \textcolor{black}{tTLG} as a more tunable platform than the twisted bilayer graphene in strongly correlated phenomena.
\end{abstract}



\maketitle

\section{Introduction}
When two monolayers stack and twist \textcolor{black}{relative} to each other in a magic angle, a set of peculiar properties \textcolor{black}{are exhibited}, for instance, correlated insulators, unconventional superconductivity, topological networks\textcolor{black}{,} and ferromagnetism\cite{cao2018correlated,cao2018unconventional,rickhaus2018transport,serlin2020intrinsic,balents2020superconductivity,sharpe2019emergent,yu2020electrically,huang2019coulomb}.
In magic angle twisted bilayer graphene, bands near the Fermi energy become ultraflat, which are believed to be responsible for most of these exotic behaviors\cite{bistritzer2011moire,zhang2019nearly,liu2019quantum}.  The realization of the flat band \textcolor{black}{can} be achieved by varying the rotation angles between two layers, trilayer rhombohedral graphene on hexagonal boron nitride and twisted few-layer graphite\cite{geim2013van,cao2020tunable,huang2019antiferromagnetically,shen_correlated_2020,lee_theory_2019,PhysRevLett.123.197702,chen2019evidence,cea2019twists,zhu2018inter}. Recent experimental and theoretical works reveal that topological flat bands \textcolor{black}{are} also present in twisted trilayer graphene (tTLG)\cite{polshyn2020electrical,chen2020electrically,shi2020tunable,ma2020topological}. The correlated states in the tTLG can be tuned by twist angles, stacking arrangements and external displacement field\cite{shi2020tunable,lei2020mirror,doi:10.1021/acs.nanolett.9b04979,jia2016strain}. This \textcolor{black}{type} of flexible controllability makes twisted trilayer graphene as a new platform to study the tunability of the correlated states in the twistronics community\cite{zhu2020twisted}.

\textcolor{black}{Ultraflat bands} in the tTLG have been predicted by utilizing the continuum model\cite{lei2020mirror,li2019electronic,PhysRevResearch.2.033150,ma2020topological}. Different from the twisted bilayer graphene case, electronic properties in tTLG depend strongly on the original stacking arrangements and on which layer is twisted\cite{lei2020mirror,li2019electronic}. For instance, a set of dispersive bands coexists with ultraflat bands at charge neutrality and the dispersion of these electronic states \textcolor{black}{is} tunable under external electric fields\cite{li2019electronic,morell2013electronic}. Furthermore, the correlated states in tTLG with asymmetric stackings are asymmetric with respect to the external electric field\cite{ma2020topological,shi2020tunable}.
These exotic properties have also been predicted using a tight-binding model with fixed nearest-neighbor intralayer hoppings\cite{morell2013electronic,lopez2020electrical}.
In all the above models, the lattice relaxations have not been taken into consideration. 
Recently, the relaxation effect has been investigated by utilizing a combination of continuum model and generalized stacking fault energy \textcolor{black}{methods}\cite{doi:10.1021/acs.nanolett.9b04979}. It shows that the relaxations renormalize the quasiparticle spectrum near the Fermi energy and provide robust energetic stability to the flat bands\cite{doi:10.1021/acs.nanolett.9b04979}. More importantly, due to the symmetry and topology differences, the \textcolor{black}{so-called} magic angles for tTLG with different stacking configurations are different\cite{ma2020topological}. The twisted monolayer-bilayer system that breaks both \textcolor{black}{twofold} rotation and mirror symmetry, \textcolor{black}{shows} correlated insulating states and ferromagnetism with an associated quantum anomalous Hall effect\cite{shi2020tunable,polshyn2020electrical,chen2020electrically}, whereas the mirror-symmetric tTLG \textcolor{black}{shows} superconducting properties\cite{park2021tunable,hao2020electric}. Current theoretical efforts are mainly focused on the investigation of the electronic band structure and of the response to an interlayer asymmetric potential, a deep understanding of the effects of tiny twist angles, lattice relaxations, and the magnetic field on the flat bands of the tTLG with different crystal symmetries is still of fundamental interest.

In this paper, we utilize a full tight-binding model to \textcolor{black}{study systematically}  the electronic properties of \textcolor{black}{tTLG} in which one of the three layers is twisted by  tiny angles relative to different stacking arrangements. Atomic relaxation is considered by treating atomic interactions using a classical potential\cite{plimpton1995fast}. Our results show that lattice relaxation is indispensable for understanding the electronic properties of tTLG with tiny twist angles. In the tTLG with mirror symmetry, \textcolor{black}{a} Dirac cone coexists with ultraflat bands near the Fermi energy. The offset energy $\Delta$E ($\Delta \mathrm E=\mathrm{E_{Dirac}}-\mathrm{E_{flat}}$, \textcolor{black}{the} energy difference between the Dirac point $\mathrm{E_{Dirac}}$ and the flat bands $\mathrm{E_{flat}}$) is tunable by the model parameters in the calculation and may vary from sample to sample in experiments due to different fabrications and encapsulations. Such energy difference is relevant in controlling the correlated phases. Moreover, for tTLG with mirror symmetry, the gap between flat bands and the other bands is robust in the presence of \textcolor{black}{a} magnetic field, which is important in the investigation of the strongly correlated properties and optical properties.

\section{Geometry and model}
\begin{figure}[h]
 	\centering
 	\includegraphics[width=\columnwidth]{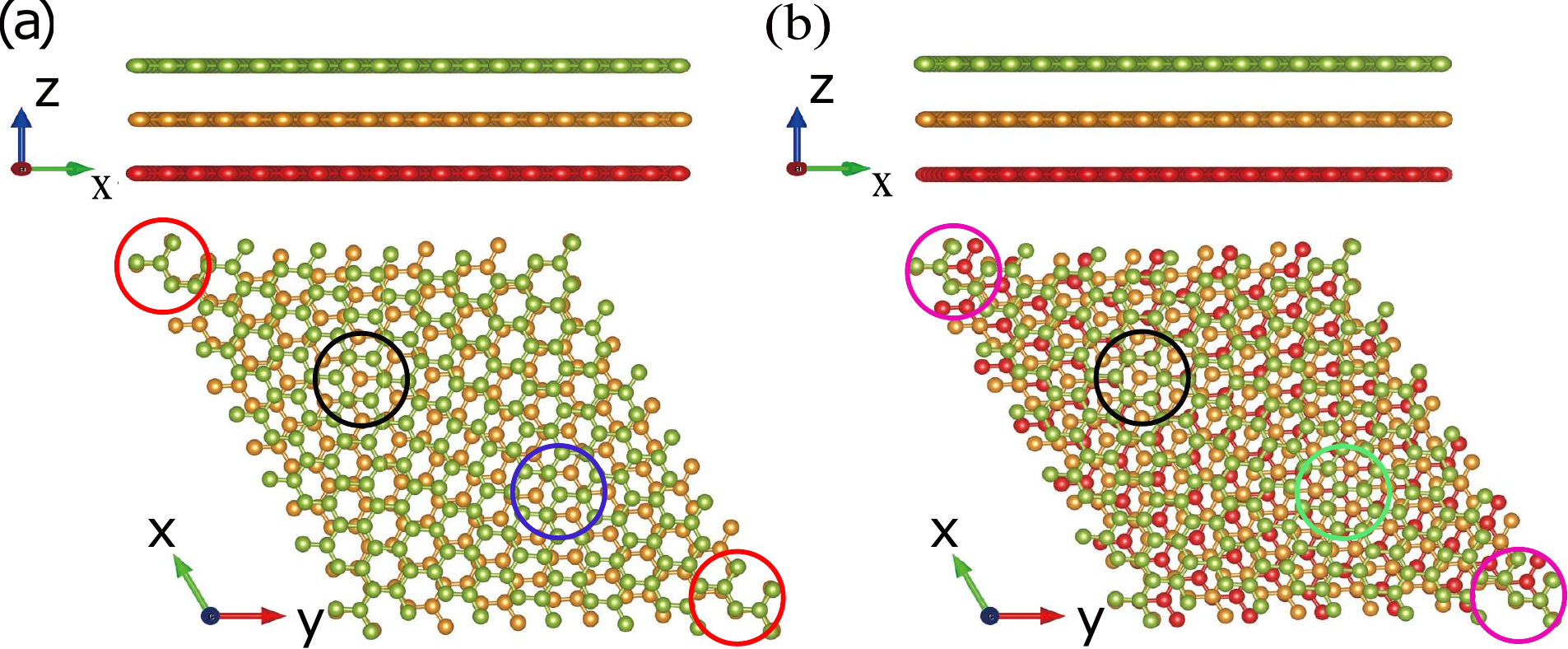}
 	\caption{The side (upper panel) and top (lower panel) views of \textcolor{black}{tTLG} (a) tTLG-A$\mathrm{\tilde{A}}$A-6.01 and (b) tTLG-$\mathrm{\tilde{A}}$AB-6.01. The high-symmetry stackings AAA, ABA, BAB, AAB, and ABC are highlighted by circles of red, black, blue, purple, and green colors, respectively. The number $6.01$ stands for the twist angle \textcolor{black}{$6.01^\circ$}.}
\label{struct}
\end{figure}
As shown in Fig. \ref{struct}, we focus on two typical twisted trilayers: one is the tTLG-A$\mathrm{\tilde{A}}$A (top and bottom layers have mirror symmetry \textcolor{black}{with} respect to the middle one and only the middle layer is twisted with a small angle), the other is the tTLG-$\mathrm{\tilde{A}}$AB (monolayer graphene on the top of AB stacking bilayer graphene and only the top layer has a relatively small twist angle). The tTLG-A$\mathrm{\tilde{A}}$A has $D_3$ symmetry and an extra mirror symmetry with the middle layer as the mirror plane, whereas the tTLG-$\mathrm{\tilde{A}}$AB has a reduced $C_3$ symmetry and breaks both \textcolor{black}{twofold} rotation and mirror symmetry. Some other stacking arrangements are discussed in the Supporting Information. We emphasize that the systems studied in this paper are different from the ``moir\'e to moir\'e'' systems, in which the electronic properties can be tuned by two independent twist angles\cite{zhu2020twisted}. Following the same methods as in twisted bilayer graphene, the tTLG can be constructed by identifying a common periodicity among the three layers\cite{shi_large-area_2020}.

The electronic properties of the \textcolor{black}{tTLG} are calculated by using a full tight-binding model based on $p_z$ orbitals. The Hamiltonian has the form:
\begin{equation}
H=\displaystyle\sum_{i}\epsilon_i|i\rangle\langle i|+\displaystyle\sum_{\langle i,j\rangle}t_{ij}|i\rangle \langle i|,
\end{equation}
where $|i\rangle$ is the $p_z$ orbital located at $\mathbf r_i$, $\epsilon_i$ is the on-site potential, and $\langle i,j\rangle$ is the sum on \textcolor{black}{indexes} $i$ and $j$ with $i\neq j$. The hopping parameter between sites $i$ and $j$ is described by a distance-related function as\cite{PhysRev.94.1498}:
 \begin{equation}
 t_{ij} = n^2 V_{pp\sigma}(r_{ij}) +(1-n^2)V_{pp\pi}(r_{ij}),
 \end{equation}
here $n$ denotes the direction cosine of $\mathbf r_{ij}=\mathbf r_j - \mathbf r_i$ along the $z$ axis and $r_{ij}=|\mathbf r_{ij}|$. The Slater and Koster parameters are defined as:
 \begin{eqnarray}
 V_{pp\sigma}(\left | r_{ij} \right |) = -\gamma_{1}e^{2.218(h-\left | r_{ij} \right |)}F_{c}(\left | r_{ij} \right |),\nonumber \\
 V_{pp\pi}(\left | r_{ij} \right |) = -\gamma_{0}e^{2.218(b-\left | r_{ij} \right |)}F_{c}(\left | r_{ij} \right |),
 \end{eqnarray}
\textcolor{black}{where} $b$ and $h$ represent nearest carbon-carbon and interlayer distances, respectively, $\gamma_0$ and $\gamma_1$ are commonly reparameterized to fit different experimental results, $F_{c}=(1 + e^{(r-0.265)/5})^{-1}$ is a smooth function. We only consider the interlayer hoppings between adjacent layers. When considering the magnetic field, the hopping $t_{ij}$ has a phase term which defined by Peierls substitution\cite{PhysRevB.82.115448,yu_dodecagonal_2019}. This full tight-binding model is accurate enough for both twisted bilayer and multilayer graphene systems\cite{shi_large-area_2020,yu_electronic_2020}.

\begin{figure*}[t]
 	\centering
 	\includegraphics[width=\textwidth]{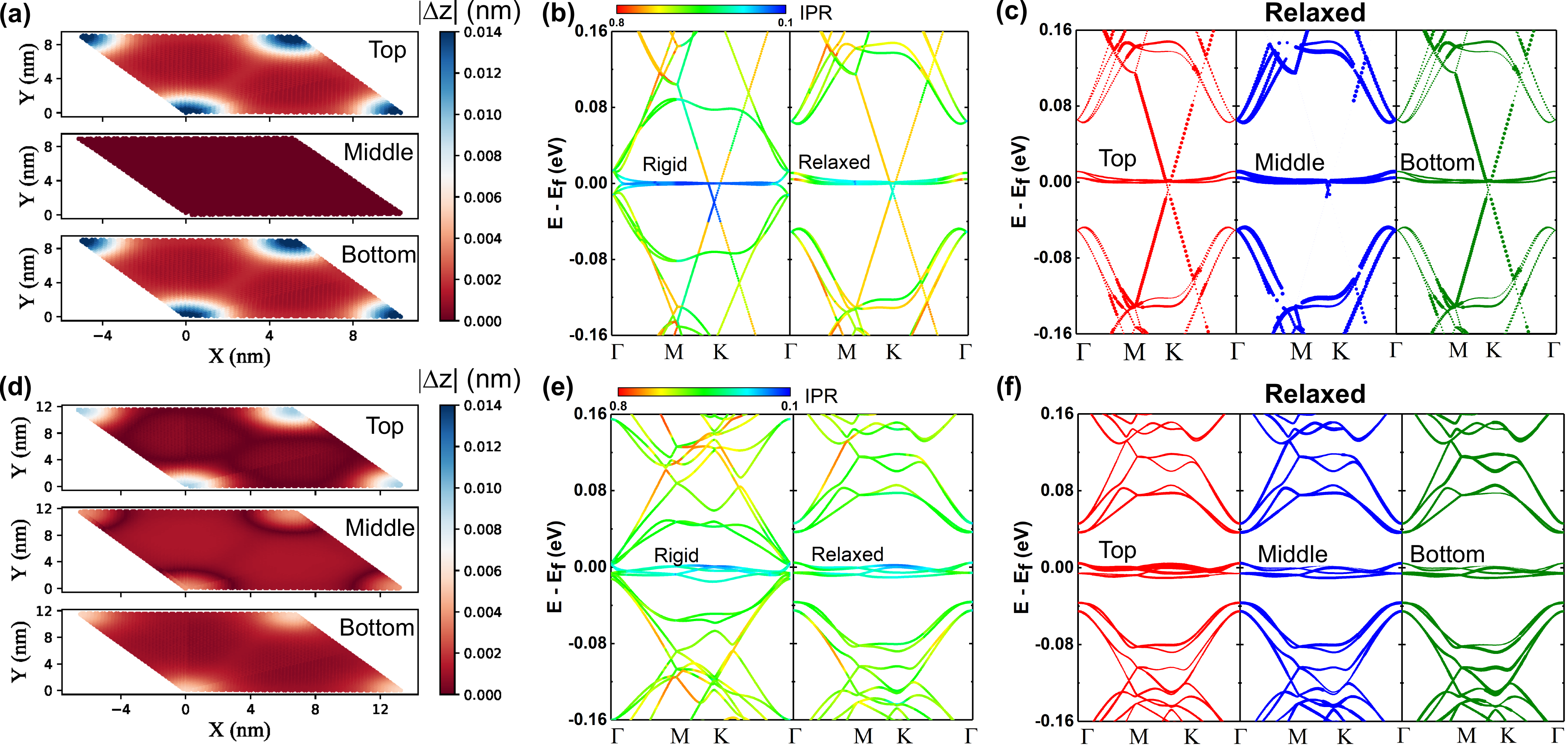}
 	\caption{Lattice relaxation effects in twisted trilayer graphene. (a) and (d) the out-of-plane displacement, (b) and (e) the band structure with the inverse participation ratio (IPR) represented by colors, (c) and (f) the layer-projected weights of band eigenstates represented by the thickness of the lines in tTLG-A$\mathrm{\tilde{A}}$A-1.35 and tTLG-$\mathrm{\tilde{A}}$AB-1.05, respectively. The hopping parameters are $\gamma_0=3.2$ eV and $\gamma_1=0.48$ eV.}
\label{relax}
 \end{figure*}

\section{Lattice relaxation effects}
We employ the classical simulation package LAMMPS to do the full relaxation\cite{plimpton1995fast}. The intralayer and interlayer interactions are simulated with LCBOP\cite{los2003intrinsic} and Kolmogorov-Crespi\cite{PhysRevB.71.235415} potentials, respectively. As illustrated in Fig. \ref{relax}(a), for the tTLG-A$\mathrm{\tilde{A}}$A, the relaxation pattern is similar to that of twisted bilayer graphene. The z component of the atom positions in the AAA region (red circle in Fig. \ref{struct}(a)) deform up to 0.14 \AA \; in both top and bottom layers , whereas the atoms in the ABA/BAB regions move \textcolor{black}{toward} \textcolor{black}{out-of-plane} only around 0.02 \AA \cite{PhysRevB.96.075311,Wijk_2015,PhysRevB.99.195419}. Differently, the atoms in the middle layer \textcolor{black}{that} sandwiches between the top and bottom layers freeze in the z-direction. The atom movements of the top and bottom \textcolor{black}{layers} have mirror symmetry \textcolor{black}{with} respect to the middle layer. The deformations are different in the tTLG-$\mathrm{\tilde{A}}$AB. In the AAB region (the purple circle in Fig. \ref{struct}(b)), the atoms on the top layer have maximum movements of 0.1 \AA. The atoms on the middle layer also deform to minimize the interlayer energy. All in all, the lattice relaxations are different for samples \textcolor{black}{in the} presence or absence of mirror symmetry.

\textcolor{black}{Mirror} symmetry also has a significant effect \textcolor{black}{on} the electronic properties of tTLG. After full relaxations, we find that the twist angle \textcolor{black}{of} tTLG-A$\mathrm{\tilde{A}}$A has the flattest band shifts from $1.47^{\circ}$ to $1.35^{\circ}$, whereas the angle in the tTLG-$\mathrm{\tilde{A}}$AB case remains the same $1.05^{\circ}$ (\textcolor{black}{for} the band structure of various twist angles see the Supporting Information). The magic angle in tTLG-A$\mathrm{\tilde{A}}$A is approximately $\sqrt{2}$ times larger \textcolor{black}{than} that in the tTLG-$\mathrm{\tilde{A}}$AB. \textcolor{black}{In general,} a larger angle corresponds to higher superconducting $\mathrm{T_C}$\cite{yankowitz2019tuning,liu2020tunable}.
Moreover, as shown in Fig. \ref{relax}(b) and (d), the lattice relaxations open a bandgap at \textcolor{black}{the} $\Gamma$ point of the Brillouin zone. The bandgap induced in tTLG-A$\mathrm{\tilde{A}}$A-1.35 is about 55.4 meV, which is two times larger than the gap in relaxed tTLG-$\mathrm{\tilde{A}}$AB-1.05 (23.4 meV). In the tTLG-A$\mathrm{\tilde{A}}$A-1.35 where mirror symmetry \textcolor{black}{is} retained, the ultraflat bands and Dirac cone still coexist.  Comparing with the rigid case, the offset energy $\Delta$E decreases from 18.8 meV to 8.4 meV, and the Fermi velocity of the Dirac cone \textcolor{black}{decreases} from $8.52*10^{5}$ m/s to $4.82*10^{5}$ m/s\cite{PhysRevB.95.085420}. Note that the Dirac cone is below the flat bands here, which is different from the previous results where the Dirac cone is above the flat bands\cite{doi:10.1021/acs.nanolett.9b04979,lopez2020electrical}. We will discuss the difference later.

Now we investigate the localization of the states in the tTLG with different stacking arrangements. This can be characterized by the inverse participation ratio (IPR), which is defined as $\sum_{i=1}^N|a_i|^2/(N\sum_{i=1}^N|a_i|^4)$, \textcolor{black}{where} $a_i$ is the state at site $i$ and $N$ is the total number of sites. Small values of the IPR correspond to localized states in the tTLG. From the results in Fig. \ref{relax}(b) and (e), it is clear that the flat bands have a higher degree of localization than the other bands. Furthermore, \textcolor{black}{lattice relaxation reduces} the localization of the flat bands. Let us focus on the IPR of tTLG-A$\mathrm{\tilde{A}}$A-1.35. The states near the Dirac cone have similar localization \textcolor{black}{to} the flat bands in both rigid and relaxed cases. However, as reported in Ref.\onlinecite{lopez2020electrical} where only constant nearest-neighbor intralayer hoppings were considered in the tight-binding model, these states have less localization than the flat bands. \textcolor{black}{Both} the localization and position of the Dirac cone are sensitive to the intralayer hoppings\cite{doi:10.1021/acs.nanolett.9b04979}. To better understand the band structure of the tTLG with different symmetries, we plot the layer projection weights of band eigenstates in Fig. \ref{relax}(c) and (f). In the relaxed tTLG-A$\mathrm{\tilde{A}}$A-1.35 with mirror symmetry, the states near the Dirac cone are only in the middle layer and the rest of the states in the dispersive bands are only in the outmost layers. The outmost layer weights are always identical. The states of the flat bands are present in all three layers and with 50\% weight in the middle layer. \textcolor{black}{In contrast}, in the relaxed tTLG-$\mathrm{\tilde{A}}$AB-1.05, the ultraflat bands have a large amount of weight from both top and middle layers, which implies an entanglement between these two layers.

\begin{figure*}[t]
 	\centering
 	\includegraphics[width=0.95\textwidth]{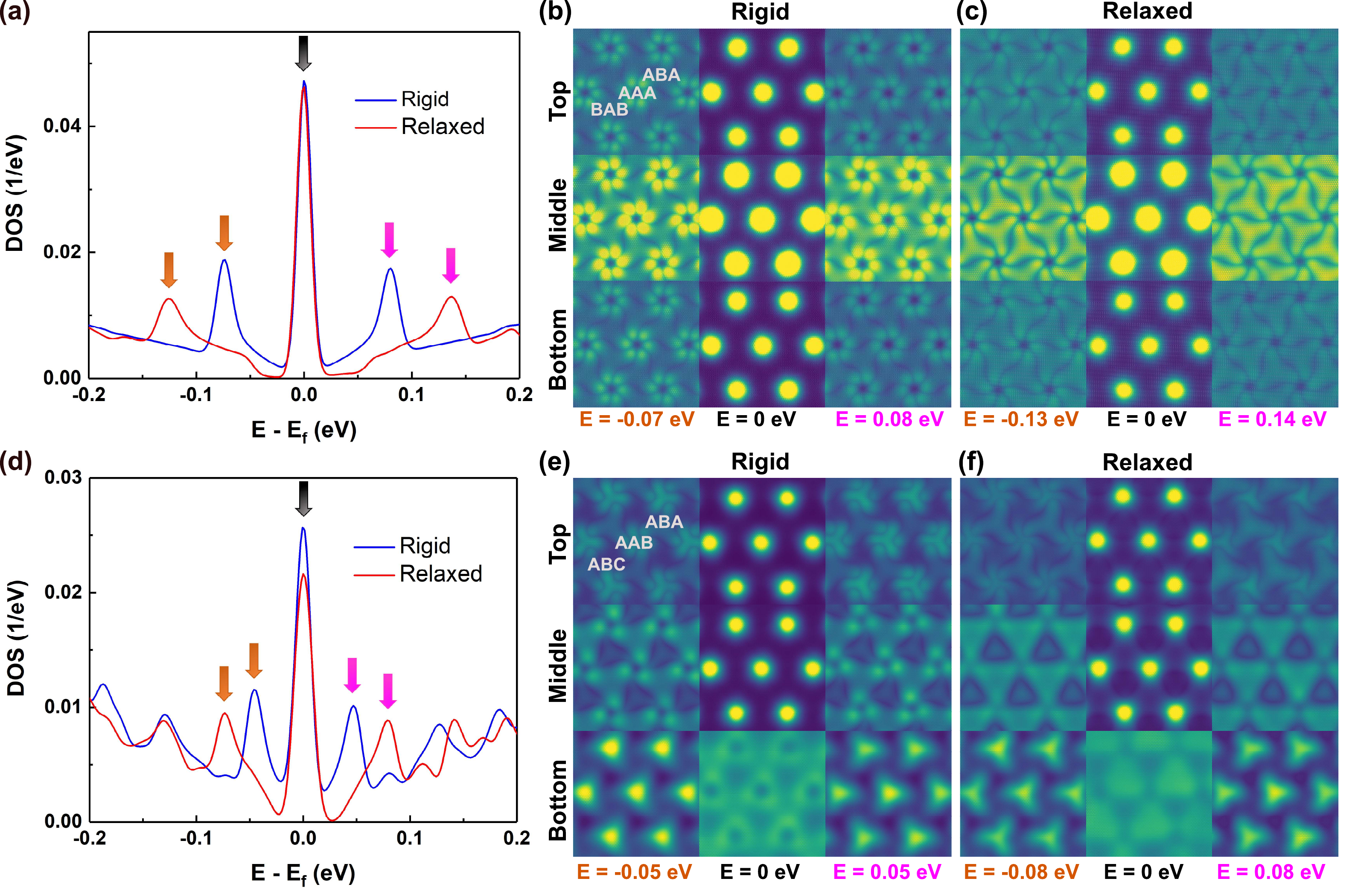}
 	\caption{Calculated local density of states \textcolor{black}{(LDOS)} mappings of different energies in the tTLG.  (a) The density of states (DOS) of tTLG-A$\mathrm{\tilde{A}}$A-1.35. (b) and (c) \textcolor{black}{LDOS} mappings of tTLG-A$\mathrm{\tilde{A}}$A-1.35 without and with lattice relaxations, respectively. Energies of the LDOS are selected near the Fermi energy in (a). (e) and (f) The LDOS mappings of tTLG-$\mathrm{\tilde{A}}$AB-1.05 without and with relaxations, respectively. Energies of the LDOS pick up from the three peaks near the Fermi energy in (d). The hopping parameters are $\gamma_0=3.2$ eV and $\gamma_1=0.48$ eV.}
 	\label{map}
\end{figure*}

We calculate the local density of states \textcolor{black}{(LDOS)} mapping to further investigate the localization of the states in real-space. As shown in Fig. \ref{map}, we focus on the states of three peaks near charge neutrality. For unrelaxed tTLG-A$\mathrm{\tilde{A}}$A-1.35 superlattice, the three different \textcolor{black}{types} of quasi-eigenstates (a superposition state of \textcolor{black}{degenerate eigenstates})\cite{PhysRevB.82.115448} in Fig. \ref{map}(b) are mainly around AAA stacking regions which is similar to the twisted bilayer graphene case\cite{shi_large-area_2020}. A large amount of flat band states localize in the middle layer, which is consistent with the layer-projected weights in Fig.\ref{relax}(c). More specifically, the neutral point states locate exactly in the AAA stacking regions whereas the other two smaller Van Hove singularity states are around the AAA regions. Moreover, the quasi-eigenstates of the top and bottom layers show mirror symmetry which could reflect consistently the mirror symmetry of the tTLG-A$\mathrm{\tilde{A}}$A structure\cite{stepanov2020untying}. After lattice relaxation, two gaps with \textcolor{black}{values} of 50 meV appear at the $\Gamma$ point near the Fermi level in Fig. \ref{map}(a). We can see two smaller Van Hove singularities move \textcolor{black}{toward a} higher energy range with its corresponding states \textcolor{black}{changing} localizations from the AAA adjacent regions to the ABA/BAB stacking regions. Similar to the twisted bilayer case, the ABA/BAB regions expand to minimize the intralayer energy. As shown in Fig. \ref{map}(b) and (c), this effect is achieved by a clockwise rotation of the moir\'e pattern in the outmost layers and a counterclockwise rotation of the moir\'e pattern in the middle layer around the AAA regions\cite{doi:10.1021/acs.nanolett.9b04979}. 

The \textcolor{black}{LDOS} mappings in the tTLG-$\mathrm{\tilde{A}}$AB-1.05 are completely different from \textcolor{black}{those} in the tTLG-A$\mathrm{\tilde{A}}$A-1.35. Comparing the results in Fig.\ref{map}(e) and (f), it is clear that the lattice relaxations have minor changes to the localization of the flat band states.  Moreover, the neutral point states are mainly localized in the top and middle layers, which indicates that the formation of flat bands in the Fermi energy is due to the interlayer interaction between the top and middle layers. The quasi-eigenstates of the other two peaks have a large part in the bottom layer in both unrelaxed and relaxed cases. Such \textcolor{black}{LDOS} mapping can be detected by a local probe such as scanning tunneling \textcolor{black}{microscopy}.

\begin{figure}[h!]
 	\centering
 	\includegraphics[width=\columnwidth]{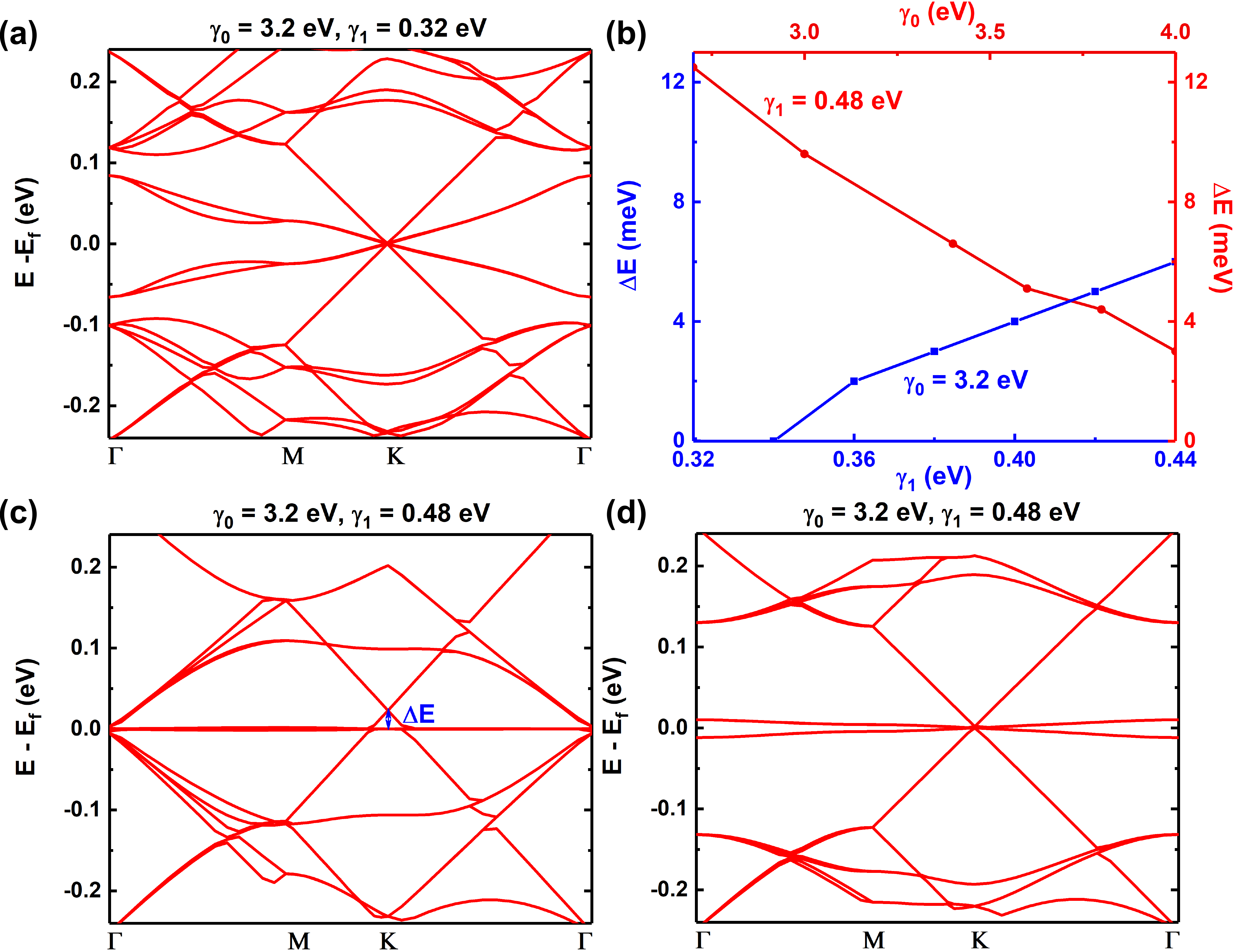}
 	\caption{Variations of the offset energy $\Delta$E in the relaxed tTLG-A$\mathrm{\tilde{A}}$A-1.35. (a) The band structure obtained with $\gamma_0=3.2$ eV and $\gamma_1=0.32$ eV. (b) The evolution of the $\Delta$E with different $\gamma_0$ (fixed $\gamma_1$) and different $\gamma_1$ (fixed $\gamma_0$) illustrated in the blue and red lines, respectively. (c) The band structure calculated with only the nearest-neighbor intralayer hoppings. (d) The band structure calculated by assuming that the interlayer hoppings of A and B sublattices between two layers are reduced to zero.}
 	\label{deltaE}
\end{figure}

\section{Shift of the Dirac cone}
A key feature in the tTLG-A$\mathrm{\tilde{A}}$A is the coexistence of flat bands with a Dirac cone in close proximity to one another in the band structure. However, in the theoretical results, some  \textcolor{black}{studies predict} the Dirac cone above the flat bands and some show it below\cite{li2019electronic,doi:10.1021/acs.nanolett.9b04979,lopez2020electrical,hao2020electric,park2021tunable}. Therefore, it is important to understand the origin of the offset energy $\Delta$E. Due to the electron hole asymmetry, the monolayer and twisted bilayer graphene  \textcolor{black}{has} an energy offset, which can be safely ignored by shifting the Fermi energy to zero after the band structure calculations are performed. However, the twisted trilayer is different. In the tTLG-A$\mathrm{\tilde{A}}$A, the flat bands have a modified Fermi energy due to the interlayer interaction and the decoupled Dirac cone preserves the monolayer energy reference, which results in a relative offset between the flat bands and the Dirac cone\cite{doi:10.1021/acs.nanolett.9b04979}.

In the monolayer, the Dirac cone is very sensitive to the model details. The parameters in the full tight-binding model can essentially influence the energy position of the Dirac cone. When the intralayer interaction is far beyond the nearest-neighbor hopping in the tight-binding model, in the calculated band structure, the Dirac cone shifts from zero to a deep negative energy and the electron-hole symmetry is broken. Similarly, comparing the band structures in Fig. \ref{deltaE}(c) and Fig. \ref{relax}(b), the Dirac cone shifts from 18.8 meV to -23.6 meV. In the above band structures, we only modify the intralayer hopping terms in the calculations. Moreover, as shown in Fig. \ref{deltaE}(a) and (b), the $\Delta$E varies with hopping parameters $\gamma_0$ and $\gamma_1$. We observe the flat bands piercing the Dirac cone when the $\gamma_1$ changes from 0.48 eV to 0.32 eV. It has been proven that the e-e interactions or extrinsic effects shift the Dirac cone \textcolor{black}{down} to the flat bands' Fermi energy\cite{doi:10.1021/acs.nanolett.9b04979}. This can be realized in the tight-binding model by reducing the interlayer hopping $\gamma_1$ between the A or B sublattice of monolayer graphene to 0. The calculated band structure is illustrated in Fig. \ref{deltaE}(d). The $\Delta$E changes from 18.8 meV to 0. From an experimental point of view, the $\Delta$E can be tuned by extrinsic factors, for instance, the way samples are prepared, the atomic force microscope-brooming procedure, and hexagonal boron nitride encapsulation. Such offset energy can be confirmed experimentally in the magnetotransport results under \textcolor{black}{a} perpendicular magnetic field. Consequently, the value of the offset energy provides information about the strength of the hopping interaction, the Fermi velocity of the Dirac cone and the many-body effects of the system. Moreover, by knowing the tendency of $\Delta E$, the relative position of the Dirac point and flat bands can be precisely controlled in  \textcolor{black}{an} experiment. Recent experiments prove that the existence of the Dirac cone allows us to control the bandwidth of the flat band, and the electrons in the Dirac bands may participate in the correlation-driven phenomena in the flat bands via the Coulomb interactions\cite{park2021tunable,hao2020electric}. It is still unclear if the $\Delta$E will affect the correlation-driven phenomena in mirror-symmetric tTLG. Moreover, by realizing the coexistence of both strongly localized and ultramobile quasiparticles in the tTLG-A$\mathrm{\tilde{A}}$A, we may observe a much higher superconducting $\mathrm{T_C}$ due to the ``steep band/flat band'' scenario of superconductivity, which is quite different from the superconductivity in the twisted bilayer graphene\cite{doi:10.1021/acs.nanolett.9b04979,alexandrov1986bipolaronic}.

\begin{figure}[t]
 	\centering
 	\includegraphics[width=\columnwidth]{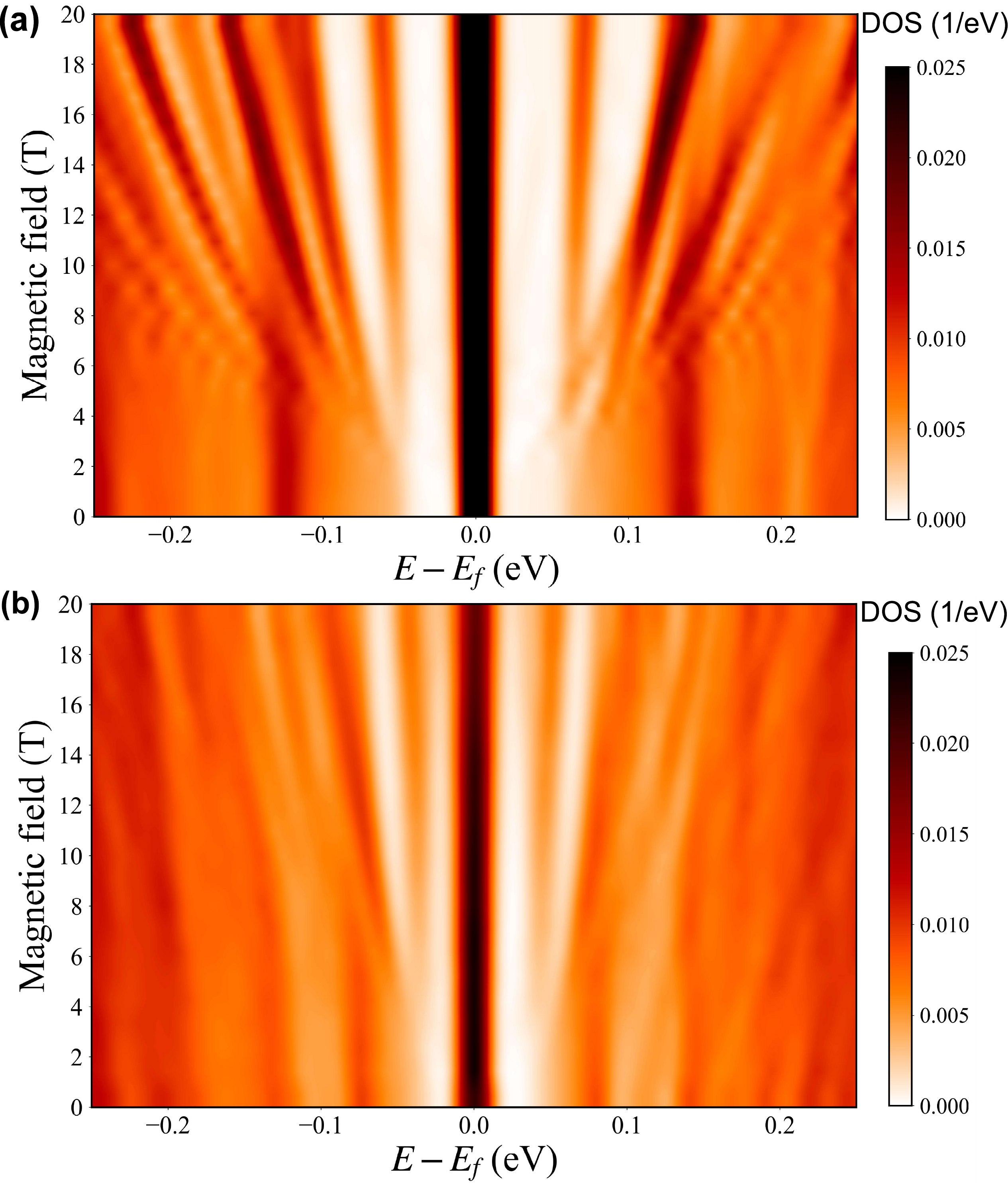}
 	\caption{(a) and (b) Hofstadter's butterflies of relaxed tTLG-A$\mathrm{\tilde{A}}$A-1.35 and tTLG-$\mathrm{\tilde{A}}$AB-1.05 with magnetic field less than 20 T, respectively. }
\label{mag}
\end{figure}
\section{The magnetic field effect}
How will the electronic states respond to an external perpendicular magnetic field in \textcolor{black}{tTLG} with or without mirror symmetry? To answer this question, we plot in Fig. \ref{mag} the Hofstadter butterfly spectrum of tTLG in the presence of strong magnetic fields. The quantized energy spectrum of tTLG in the magnetic field \textcolor{black}{is quite} different for different stackings. In the tTLG-A$\mathrm{\tilde{A}}$A-1.35, the Hofstadter butterfly spectra are apparent at a small field of 4 T. We can also see an obvious electron-hole asymmetry of the states \textcolor{black}{responding} to the external magnetic field. Compared with the rigid cases shown in the Supporting Information, two gaps flanking the central peak are obvious in Fig. \ref{mag}. The Landau levels (LLs) in the tTLG-$\mathrm{\tilde{A}}$AB-1.05 are characterized by a more complicated fractal energy spectrum. It needs a stronger magnetic field to realize apparent Hofstadter butterfly spectra. This is ascribed to different interlayer interactions in tTLG with different stacking arrangements. Previous results show that the interlayer coupling plays a crucial role in the formation of Hofstadter butterfly spectra\cite{wang2012fractal}. From the results in Figs. \ref{relax} and \ref{map}. it is obvious that the interlayer interaction in tTLG-A$\mathrm{\tilde{A}}$A is much stronger than that in tTLG-$\mathrm{\tilde{A}}$AB. The flatness of the low energy bands and the localization of these states efficiently affect the properties of the LLs in the presence of \textcolor{black}{a} magnetic field. Furthermore, the magnitude of the magnetic field for the occurrence of the Hofstadter butterfly spectra in tTLG is much smaller than the reported value in twisted bilayer graphene\cite{wang2012fractal}. In twisted bilayer graphene near the magic angle, the higher indices of the LL start to appear at a field of 30 T. A notable feature is that the Hofstadter butterfly spectrum is an example to exhibit fractal properties in the tTLG. It could be observed at laboratory accessible magnetic field strengths.

\section{CONCLUSION}
We systematically investigated the ultraflat bands in \textcolor{black}{tTLG} with different stacking arrangements. We found that the \textcolor{black}{tTLG} with or without  mirror symmetry \textcolor{black}{has a quite} different response to lattice relaxations and \textcolor{black}{a} magnetic field. In particular, for mirror-symmetric tTLG-A$\mathrm{\tilde{A}}$A, the electronic properties are significantly changed when the lattice relaxations are \textcolor{black}{considered}, and the gap between the flat bands and the other bands \textcolor{black}{is} robust in the presence of magnetic fields. Moreover, the energy offset between the flat bands and the Dirac cone vertex is sensitive to the parametrization of the model, which corresponds to the extrinsic factors introduced during the experiment. In the tTLG-$\mathrm{\tilde{A}}$AB with a reduced crystal symmetry, the Dirac cone disappears when the twist angle is tiny, and the lattice relaxations have less influence \textcolor{black}{on} its electronic properties. Furthermore, the flat bands are robust under \textcolor{black}{a} laboratory reachable magnetic field. Although we do not calculate the correlation strengths directly, we can use the tunable flat bands \textcolor{black}{at} charge neutrality as a proxy for the electronic correlation, and provide a starting point to explore the interplay between the flat bands, mirror symmetry, magnetic field, dispersive states and correlation in \textcolor{black}{tTLG}.



~\\
\textbf{Acknowledgments}

This work was supported by the National Natural Science Foundation of China (Grants No.11774269 and No.12047543), the National Key R\&D Program of China (Grant No. 2018FYA0305800), and the Natural Science Foundation of Hubei Province, China (2020CFA041). Numerical calculations presented in this paper \textcolor{black}{were} performed on the supercomputing system in the Supercomputing Center of Wuhan University.

%

\bibliography{references.bib}

\end{document}



\title{Lattice relaxation, mirror symmetry and magnetic field effects on ultraflat bands in twisted trilayer graphene}

\author{Zewen Wu}
\affiliation{Key Laboratory of Artificial Micro- and Nano-structures of Ministry of Education and School of Physics and Technology, Wuhan University, Wuhan 430072, China}

\author{Zhen Zhan}
\email{zhen.zhan@whu.edu.cn}
\affiliation{Key Laboratory of Artificial Micro- and Nano-structures of Ministry of Education and School of Physics and Technology, Wuhan University, Wuhan 430072, China}

\author{Shengjun Yuan}
\email{s.yuan@whu.edu.cn}
\affiliation{Key Laboratory of Artificial Micro- and Nano-structures of Ministry of Education and School of Physics and Technology, Wuhan University, Wuhan 430072, China}


\date{\today}

\begin{abstract}

\end{abstract}

\maketitle
\subsection{The band structure of twisted trilayer graphene with different stackings}
\begin{figure}[h]
	\centering
	\includegraphics[width=\columnwidth]{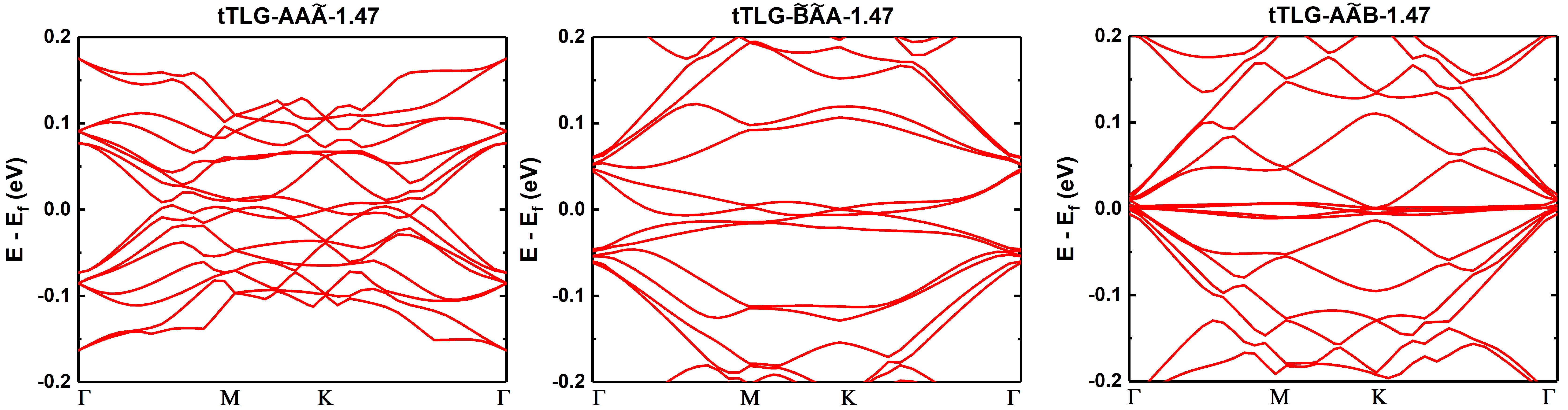}
	\caption{The band structure of rigidly twisted trilayer graphene with the twist angle $\theta=1.47^\circ$. (a) AAA-stacked with a bottom layer twist. (b) BAA-stacked with top and middle layers twist. (c) AAB-stacked with a middle layer twist. In the tTLG-AA$\mathrm{\tilde{A}}$-1.47 sample, the starting stacking arrangement is AAA, the twist layer is marked by a tilde above the letter, and the last number represents the rotation angle.}
	\label{stackings}
\end{figure}

We investigate the band structure of twisted trilayer graphene (tTLG) with different stacking arrangements. The moir\'e supercells investigated in Fig. \ref{stackings} are mirror asymmetric. For tTLG with an outmost-layer twist, for instance, tTLG-AA$\mathrm{\tilde{A}}$, we find a Dirac cone at the K point of the Brillouin zones (BZ), which has a reduced Fermi velocity. Flat bands and vanishing velocities are not found in this stacking. The tTLG-$\mathrm{\tilde{B}}$$\mathrm{\tilde{A}}$A and tTLG-$\mathrm{\tilde{A}}$AB (results are shown in the main text) have the same starting stacking arrangements but different twist layer, which results in minor difference in the band structure. In tTLG-A$\mathrm{\tilde{A}}$B, we find flat bands near the charge neutrality in company with a parabolic band at the K point. In the above results, the only difference is the starting stacking arrangements. It is obvious that the stacking arrangements change significantly the electronic properties in twisted trilayer graphene.

\subsection{The band structure of twisted trilayer graphene with different rotation angles}
\begin{figure}[h]
	\centering
	\includegraphics[width=0.9\columnwidth]{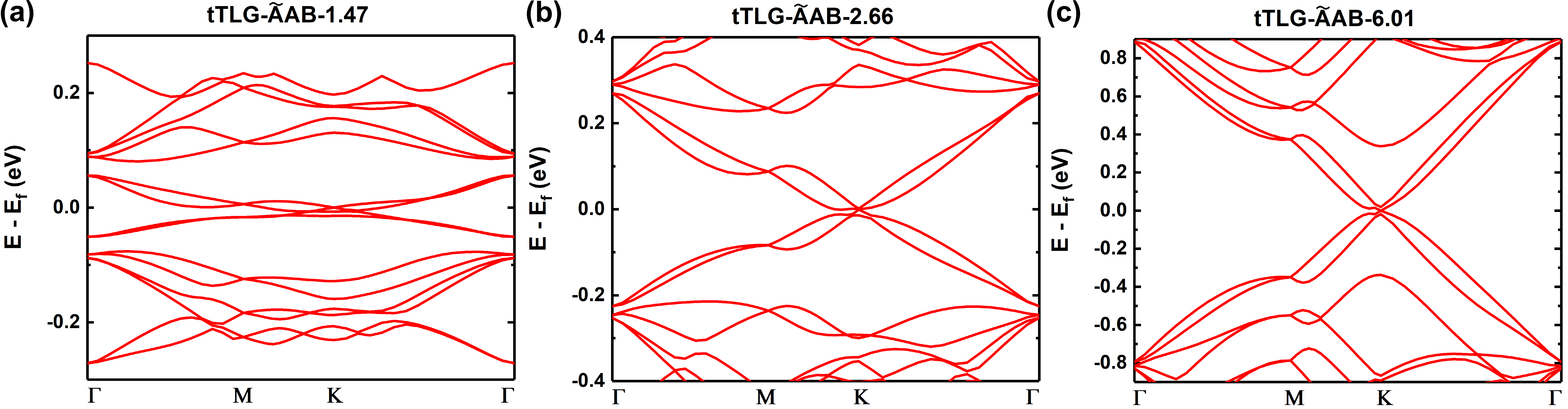}
	\caption{The band structure of relaxed tTLG-$\mathrm{\tilde{A}}$AB with twist angles (a) $\theta=1.47^\circ$, (b) $\theta=2.66^\circ$ and (c) $\theta=6.01^\circ$.}
	\label{aab_bs}
\end{figure}
\begin{figure}[h]
	\centering
	\includegraphics[width=0.9\columnwidth]{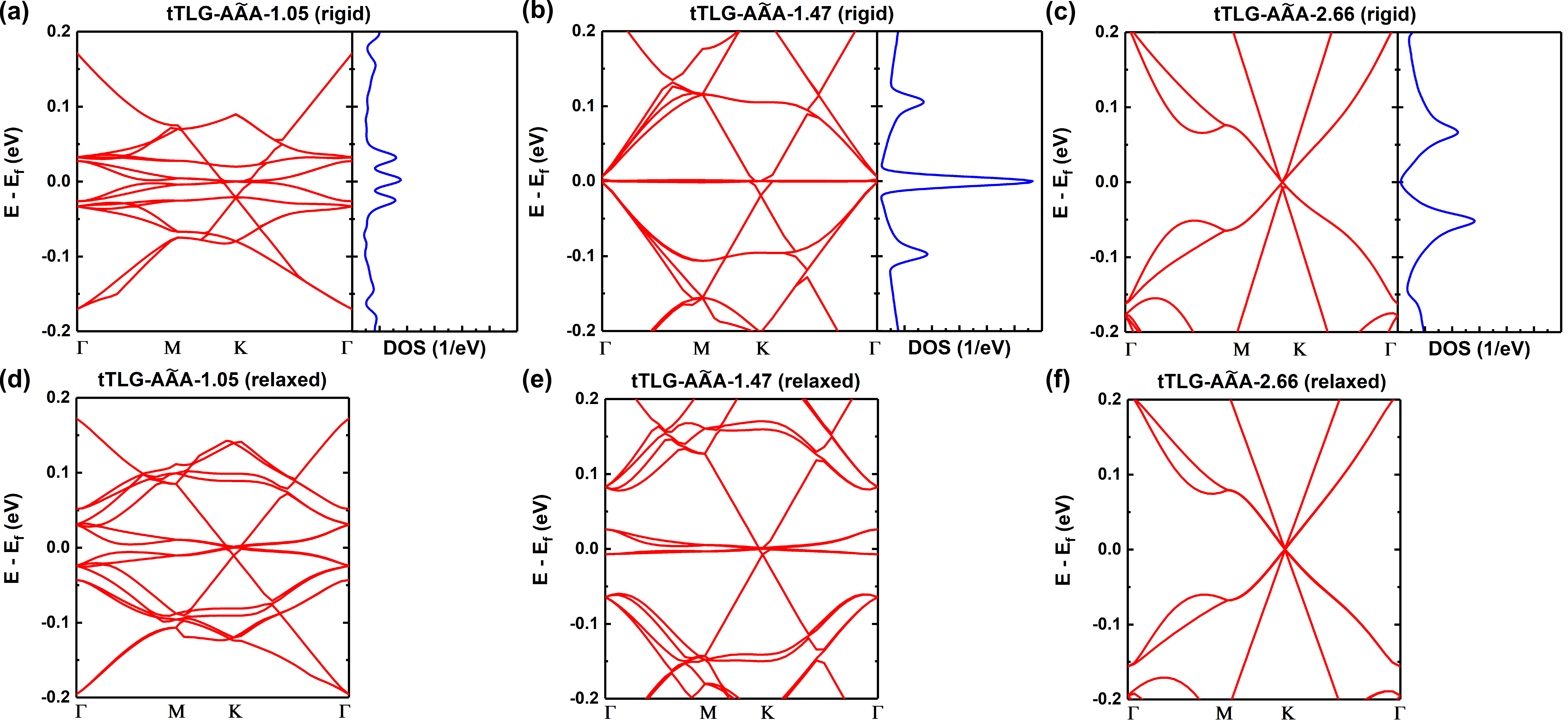}
	\caption{The band structure (red line) and density of states (blue line) of rigid tTLG-A$\mathrm{\tilde{A}}$A with twist angles (a) $\theta=1.05^\circ$, (b) $\theta=1.47^\circ$ and (c) $\theta=2.66^\circ$. (d)-(f) The band structure of relaxed tTLG-A$\mathrm{\tilde{A}}$A with $\theta=1.05^\circ$, $\theta=1.47^\circ$ and $\theta=2.66^\circ$, respectively.}
	\label{aaa_bs}
\end{figure}

In Fig. \ref{aab_bs}, we study the effect of the rotation angle on the band structure of relaxed tTLG-$\mathrm{\tilde{A}}$AB. We find two parabolic Bernal-like bands and a twisted-like Dirac cone at the K point in tTLG-$\mathrm{\tilde{A}}$AB with a large twist angle. The Dirac bands has a Fermi velocity reduction when the twist angle decreases. The lattice relaxations have non-neglectful effect to the band structures at small twist angles. A band gap is opened at the $\Gamma$ point between the four low-energy bands and the other high energy bands. The flattest bands appear at $\theta=1.05^\circ$. Comparing the Fig. \ref{aab_bs} and Fig. 2(e) of the main text, we show that the maximum band gap with a value of 23.4 meV appears in tTLG-$\mathrm{\tilde{A}}$AB-1.05.

As shown in Fig. \ref{aaa_bs}, the magic angle is around $\theta=1.47^\circ$ in rigid tTLG-A$\mathrm{\tilde{A}}$A, whereas the magic angle moves to $\theta=1.35^\circ$ in the relaxed case. Moreover, the magnitude of the density of states in magic angle supercell has the maximum value. Similar to the tTLG-$\mathrm{\tilde{A}}$AB case, a band gap opens between the flat bands and the excited bands with the maximum value around 55.4 meV at $\theta=1.35^\circ$.
\subsection{The effect of the electric field}

 \begin{figure}[h]
	\centering
	\includegraphics[width=0.8\columnwidth]{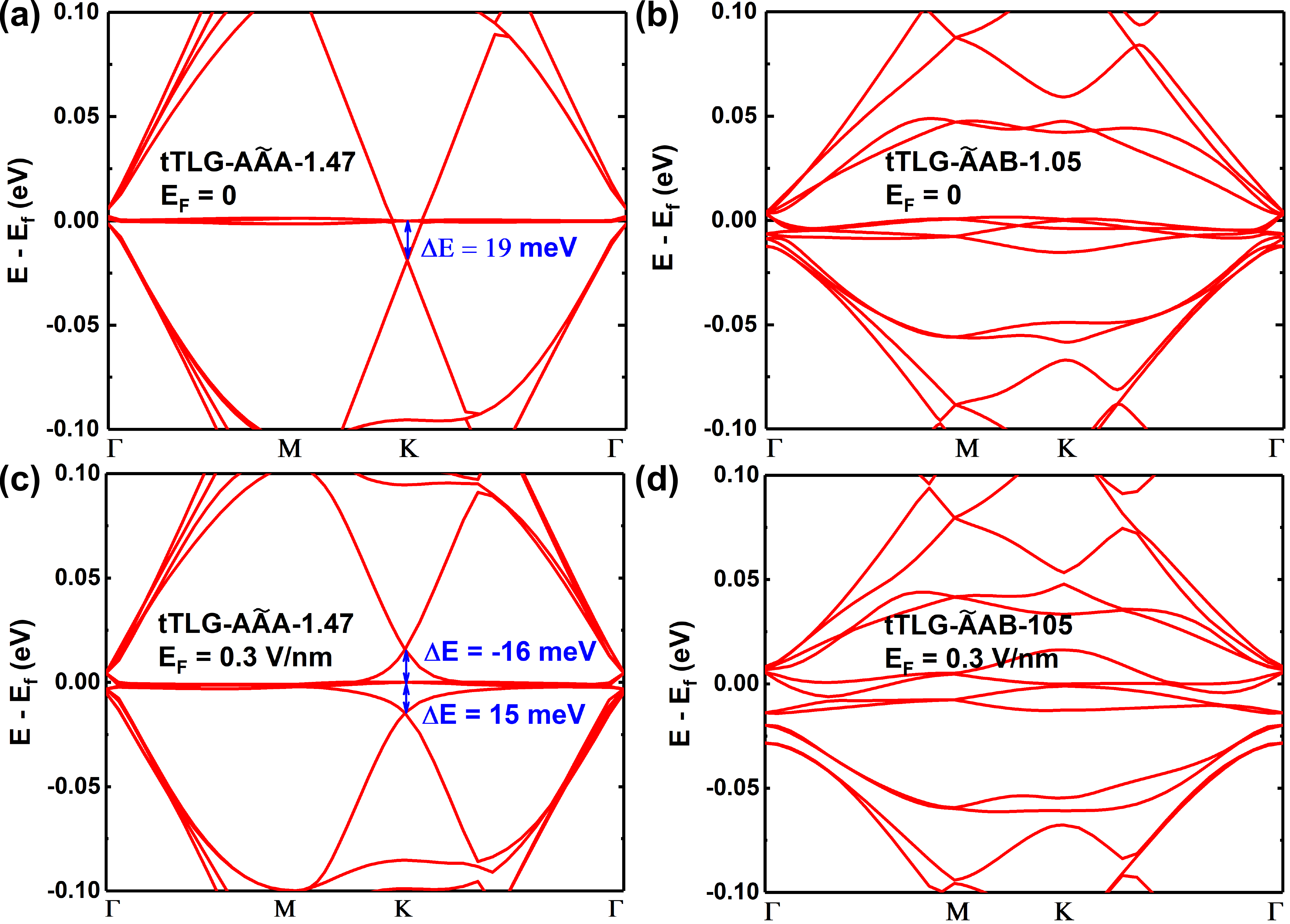}
	\caption{Band structure of rigid tTLG-A$\mathrm{\tilde{A}}$A-1.47 with external electric field (a) $\mathrm{E_F}=0$ and (b) $\mathrm{E_F}=0.3$ V/nm. (c)  and (d) Band structure of rigid tTLG-$\mathrm{\tilde{A}}$AB-1.05 with $\mathrm{E_F}=0$ and $\mathrm{E_F}=0.3$ V/nm, respectively.}
	\label{electric}
\end{figure}

As shown in Fig. \ref{electric}(a), our calculation also indicates that the band structure of tTLG-A$\mathrm{\tilde{A}}$A-1.47 shows interesting flat band and Dirac point coexist phenomenon\cite{li2019electronic,doi:10.1021/acs.nanolett.9b04979,PhysRevB.100.085109}. One interesting finding is that the Dirac point could be split into two by applying a perpendicular electric field. When an electric field results bias potential between top and bottom layers up to 200 meV, as shown in Fig. \ref{electric}(c), a new emerged Dirac point and the original one show a mirror symmetry respect to the flat bands\cite{doi:10.1021/acs.nanolett.9b04979}.
The Fermi velocity near the two Dirac points are highly suppressed to $5.078*10^{5}$ m/s (Dirac cone above the flat bands) and $5.017*10^{5}$ m/s (Dirac cone below the flat bands) compared with the static Dirac point with Fermi velocity $9.460*10^{5}$ m/s. It has been proven that the existence of the Dirac cone allows us to control the bandwidth and interaction strength in the flat bands of tTLG-A$\mathrm{\tilde{A}}$A by varying the electric field $\mathrm{E_F}$\cite{park2021tunable,hao2020electric}. In tTLG-$\mathrm{\tilde{A}}$AB-1.05, the electric field makes the bands near the charge neutrality less flat. A gap is opened between the flat bands and the other valence bands at the $\Gamma$ point of the Brillouin zone.

\subsection{The effect of the magnetic field on the electronic properies of rigidly twisted tralayer graphene}
\begin{figure}[h]
	\centering
	\includegraphics[width=0.8\columnwidth]{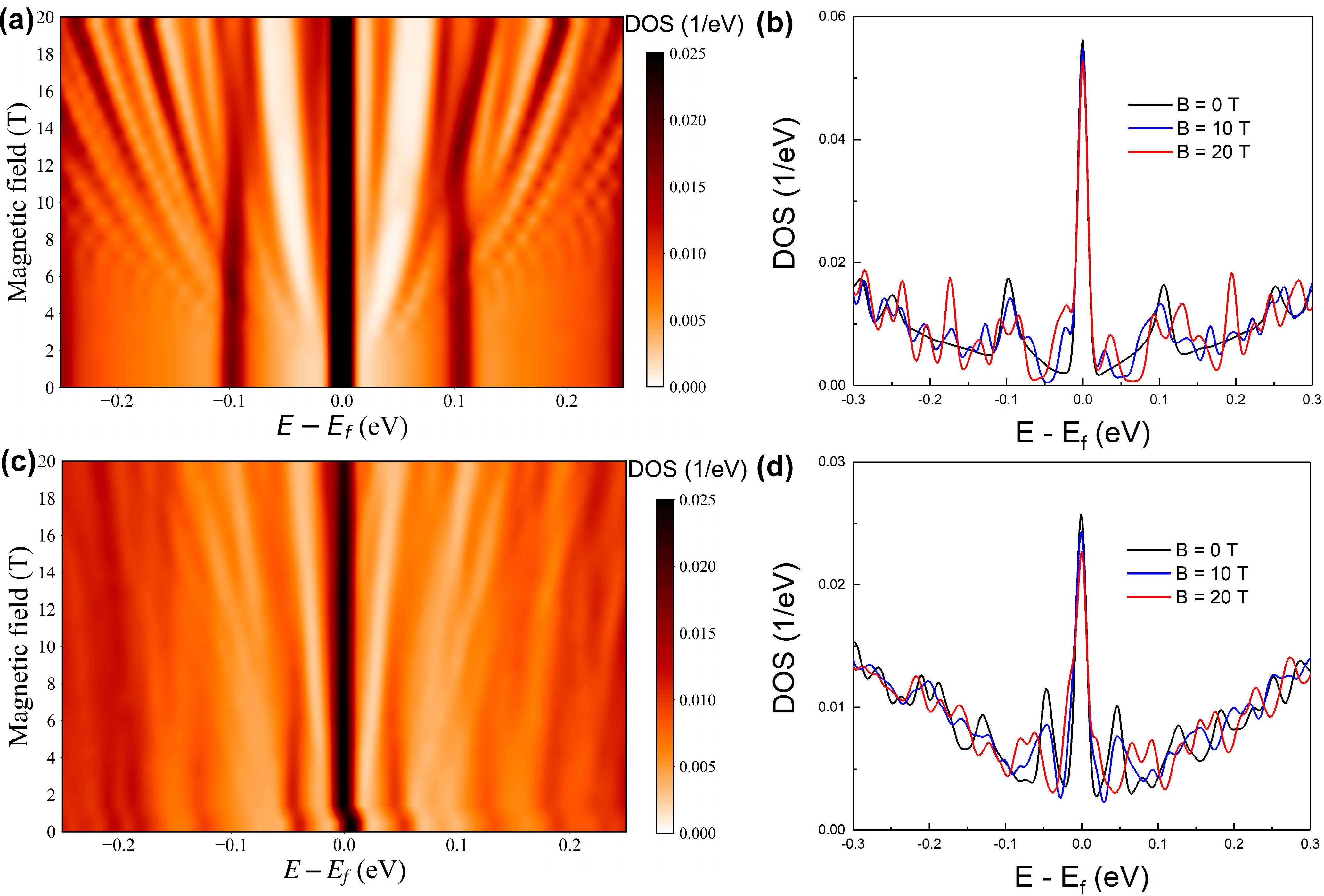}
	\caption{(a) and (c) Hofstader's butterflies of rigid tTLG-A$\mathrm{\tilde{A}}$A-1.47 and tTLG-$\mathrm{\tilde{A}}$AB-1.05 with magnetic field less than 20 T, respectively. (b) and (d) The density of states of rigid tTLG-A$\mathrm{\tilde{A}}$A-1.47 and tTLG-$\mathrm{\tilde{A}}$AB-1.05 under 0 T,10 T, and 20 T magnetic field, respectively.}
	\label{magnetic}
\end{figure}

In this part, we investigate how will the flat bands in rigid tTLG response to the external magnetic field. From the Fig. \ref{magnetic}, we can see higher Landau levels when the magnetic field under the 4 T in tTLG-A$\mathrm{\tilde{A}}$A-1.47 while for the tTLG-$\mathrm{\tilde{A}}$AB-1.05 the Landau levels with higher index occur under the 10 T magnetic field. Similar to the relaxed case in the main text, the Landau levels in mirror-asymmetry tTLG-$\mathrm{\tilde{A}}$AB are characterized by a more complicated fractal energy spectrum. Comapring with the relaxed case, there are two main differences. First, the Dirac cone can be detected in Fig. \ref{magnetic}(a). Some states appear at the left hand of the flat band peak, which are correspond to the states near the Dirac cone. Second, for both tTLG-A$\mathrm{\tilde{A}}$A-1.47 and tTLG-$\mathrm{\tilde{A}}$AB-1.05, no band gap between the flat bands and the other high energy bands is detected in the Hofstader's butterflies. However, such band gaps are found in the relaxed cases, which is essential to realize the correlated and topological states in twisted graphene multilayers.

\bibliography{references_SI.bib}